\documentclass[aps,prl,amsmath,amssymb,floatfix,twocolumn,amsmath,superscriptaddress,twocolumn,nofootinbib,tighten,letterpaper]{revtex4}
\usepackage{multirow}
\usepackage{bbold}
\usepackage{subfigure}
\usepackage{color}
\usepackage{mathrsfs}
\usepackage{hyperref}
\usepackage[normalem]{ulem}
\usepackage{bm}

\usepackage{amsfonts, relsize, color}
\usepackage{graphics}
\usepackage{graphicx}
\usepackage{subfigure}
\usepackage{hyperref}
\usepackage{color}

\begin{document}
\title{Strain engineered higher order topological phases for spin-3/2 Luttinger fermions}

\author{Andr$\acute{\mbox{a}}$s L. Szab$\acute{\mbox{o}}$}
\affiliation{Max-Planck-Institut f\"{u}r Physik komplexer Systeme, N\"{o}thnitzer Str. 38, 01187 Dresden, Germany}

\author{Roderich Moessner}
\affiliation{Max-Planck-Institut f\"{u}r Physik komplexer Systeme, N\"{o}thnitzer Str. 38, 01187 Dresden, Germany}

\author{Bitan Roy}\email{bitan.roy@lehigh.edu}
\affiliation{Max-Planck-Institut f\"{u}r Physik komplexer Systeme, N\"{o}thnitzer Str. 38, 01187 Dresden, Germany}
\affiliation{Department of Physics, Lehigh University, Bethlehem, Pennsylvania, 18015, USA}

\date{\today}
\begin{abstract}
Recently, the notion of topological phases of matter has been extended to higher-order incarnations, supporting gapless modes on even lower dimensional boundaries, such as corners and hinges. We here identify a collection of cubic spin-3/2 fermions with biquadratic touching of Kramers degenerate valence and conduction bands as a platform to strain-engineer higher-order topological (HOT) phases: external uniaxial strain gives birth to a HOT Dirac semimetal or an insulator, depending on its sign, featuring topological \emph{hinge} modes in the strain direction. The insulator in fact exhibits \emph{mixed} topology, and in addition supports edge modes on orthogonal planes. These outcomes are germane when the external strain is applied along one of the $C_{4v}$ or coordinate axes, as well as $C_{3v}$ or body-diagonal, directions. Our findings place HgTe, gray-Sn, 227 pyrochlore iridates and half-Heusler compounds at the frontier of strain-engineered electronic HOT phases.    
\end{abstract}

\maketitle

\emph{Introduction}. Strain-engineering has been of intense interest recently, being instrumental for the identification of nematic phases in correlated materials, such as cuprates, pnictides and heavy-fermion compounds~\cite{Mackenzie, pnictide-strain}. This is so, because strain couples \emph{linearly} to nematicity, similar to how an external magnetic field couples to magnetization. It can also give rise to other exotic phenomena, such as topologically protected axial Landau levels in strained Dirac and Weyl materials~\cite{manoharan-review,roy-assaad-herbut, grushin-vishwanath, pikulin-chen-Franz}, and a quantum phase transition between a topological Dirac semimetal (TDSM) and trivial band insulator~\cite{Montambaux, Esslinger, Roy-Foster,Sur-Roy}, to name a few. We propose the three-dimensional Luttinger semimetal (LSM) of spin-3/2 fermions, displaying a biquadratic touching of Kramers degenerate valence and conduction bands~\cite{luttinger, murakami-nagaosa-zhang}, as a platform to strain-engineer a plethora of higher-order topological (HOT) phases, featuring one-dimensional hinge modes of codimension $d_c=2$.

Typically, a $d$-dimensional topological state supports gapless modes on a $(d-1)$ dimensional boundary (characterized by $d_c=1$)~\cite{hasan-kane-review2010, qi-zhang-review2011,chiu-review2016, armitage-RMP2018}. This notion has been extended recently to incorporate higher order cousins~\cite{benalcazar2017,schindler2018, song2017, benalcazar-prb2017, langbehn2017, schindler-sciadv2018,ezawa2018, khalaf2018,hsu2018,trifunovic2019,wang1-2018,yan2018,calugaru2019,ahn2018,Vliu2018,rodriguez2018,franca2018, matsugatani2018, vanmiert2018, wang-arxiv2018, ghorashi2019,Klinovaja2019, agarwala2019, Tnag2019, Klinovaja2019arXiv, ZYan2019, kaisun2019, roy-singleauthor2019, Vliu2020}: an $n$th order topological phase accommodates gapless modes on a boundary of codimension $d_c=n>1$. Prototypical examples include corner and hinge states with $d_c=d$ and $d_c=d-1$, respectively. In this nomenclature, conventional topological phases are first order. While HOT phases have been realized in various metamaterials, such as phononic~\cite{serra-garcia2018} and photonic~\cite{noh2018, peterson} crystals, and electric circuits~\cite{imhof2018}, so far elemental Bi stands as the lone example to support an electronic HOT state~\cite{schindler2018}.

%%%%%%%%%%%%%%%%%%%%%%%%%%%%%%%%%%%%%%%%%%%%%%%%%%%%%
%%%%%%%%%%%%%%%%%%%%%%%%%%%%%%%%%%%%%%%%%%%%%%%%%%%%%
%%%%%%%%%%%%%%%%%%%%%%%%%%%%%%%%%%%%%%%%%%%%%%%%%%%%%
%%%%%%%%%%%%%%%%%%%%%%%%%%%%%%%%%%%%%%%%%%%%%%%%%%%%%
%%%%%%%%%%%%%%%%%%%%%%%%%%%%%%%%%%%%%%%%%%%%%%%%%%%%%
\begin{figure}[t!]
\includegraphics[width=0.98\linewidth]{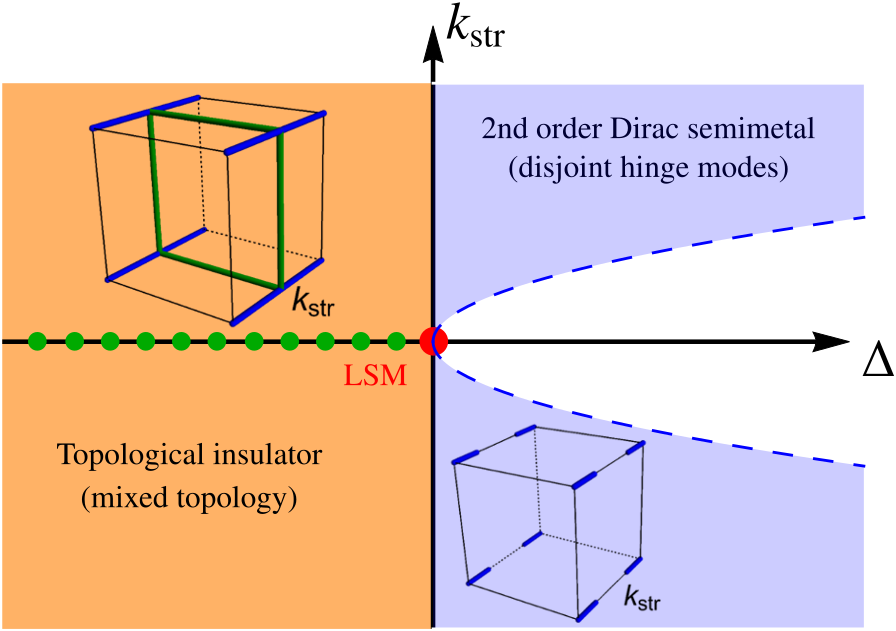}
\caption{Schematic phase diagram of a LSM (red dot), in the presence of an external strain of magnitude $|\Delta|$. Here, ${\rm sgn}(\Delta)=+ \; (-)$ corresponds to tensile (compressive) strain, which gives rise to a second order Dirac semimetal (topological insulator). The momentum in the direction of the external strain is $k_{\rm str}$. In the semimetal phase two Dirac points, located at $\pm k^{\ast}_{\rm str}$, are separated along this axis, where $k^{\ast}_{\rm str} \sim \sqrt{|\Delta|}$. Both phases produce four hinge states of codimension $d_c=2$, along $k_{\rm str}$ [see blue lines in the cubes (inset)]. In the insulating (Dirac semimetal) phase the hinge modes exist for any $k_{\rm str}$ ($|k_{\rm str}|> k^\ast_{\rm str}$). The topological insulator also supports edge modes (green lines) with $d_c=1$ on the $k_{\rm str}=0$ plane (green dots), occupying two planes orthogonal to $k_{\rm str}$; thus accommodating mixed topology. The white region ($|k_{\rm str}|<k^\ast_{\rm str}$) is devoid of any boundary mode. In practice, $k_{\rm str}$ is a high symmetry ($\langle 100 \rangle$ or $\langle 111 \rangle$, for example) direction. 
}~\label{Fig:Phasediagram}
\end{figure}
%%%%%%%%%%%%%%%%%%%%%%%%%%%%%%%%%%%%%%%%%%%%%%%%%%%%%
%%%%%%%%%%%%%%%%%%%%%%%%%%%%%%%%%%%%%%%%%%%%%%%%%%%%%
%%%%%%%%%%%%%%%%%%%%%%%%%%%%%%%%%%%%%%%%%%%%%%%%%%%%%
%%%%%%%%%%%%%%%%%%%%%%%%%%%%%%%%%%%%%%%%%%%%%%%%%%%%%
%%%%%%%%%%%%%%%%%%%%%%%%%%%%%%%%%%%%%%%%%%%%%%%%%%%%%

We here show that strain in a three-dimensional LSM can induce a variety of HOT phases, both gapless and insulating. When stress is applied to a LSM along certain high-symmetry directions, such as the $C_{4v}$ ($x,y$ and $z$ coordinate axes) and $C_{3v}$ (four body diagonal or $\langle 111 \rangle$) directions (respectively classified as the $E_g$ and $T_{2g}$ strains, according to the cubic or $O_h$ point group~\cite{goswami-roy-dassarma}), the resulting strain induces either a TDSM or a topological insulator, depending on the sign ($\pm$) of the strain (which we refer to as tensile or compressive, respectively). Both support four copies of one-dimensional hinge modes in the direction of the applied strain. In addition, the insulator supports edge modes that reside on two-dimensional surfaces ($d_c=1$) perpendicular to the applied strain, thus exhibiting \emph{mixed} topology. The resulting phase diagram is schematically shown in Fig.~\ref{Fig:Phasediagram}. Our results may be of experimental relevance to HgTe~\cite{hgte}, gray-Sn~\cite{gray-sn-1, gray-sn-2}, 227 pyrochlore iridates~\cite{Savrasov, Exp:Nakatsuji-1, Exp:Nakatsuji-2} and half-Heuslers~\cite{Exp:cava, Exp:felser, binghai}.

\emph{A toy model}. To set the stage, we first consider a model describing three-dimensional HOT phases (both insulator and semimetal), constituted by spin-$\frac{1}{2}$ fermions. The corresponding Hamiltonian can be decomposed as $\hat{h}^{\rm 3D}_{\rm HOT} ({\bf k}) =\hat{h}^{\bf k}_{0}+\hat{h}^{\bf k}_{1}$, where 
\begin{eqnarray}~\label{Eq:LatticeHamilToy}
\hat{h}^{\bf k}_{0} &=& t \sum^2_{j=1} S_j \gamma_j + \bigg[ t_z C_z -m_z+ t_0 \sum^2_{j=1}(1-C_j) \bigg] \gamma_3, \nonumber \\
\hat{h}^{\bf k}_{1} &=& \Delta_1 (C_2-C_1) \gamma_4 + \Delta_2 S_1 S_2 \gamma_5, 
\end{eqnarray} 
$C_j \equiv \cos(k_j a)$, $S_j \equiv \sin(k_j a)$, $k_j$s are components of momenta, and $a$ is the lattice spacing, set to unity. Four-component $\gamma$ matrices satisfy the anticommuting Clifford algebra $\{\gamma_j, \gamma_k\}=2 \delta_{jk}$, for $j,k=1,\cdots,5$. Even though the following discussion does not depend on the representation of $\gamma$ matrices, for the sake of concreteness we choose $\gamma_1=\sigma_3 \tau_1$, $\gamma_2=\sigma_0 \tau_2$, $\gamma_3=\sigma_0 \tau_3$, $\gamma_4=\sigma_1 \tau_1$, and $\gamma_5=\sigma_2 \tau_1$. Two sets of Pauli matrices $\{ \sigma_\mu \}$ and $\{ \tau_\mu \}$ respectively operate on spin and sublattice/orbital indices, where $\mu=0,1,2,3$.

Expanding $\hat{h}^{\rm 3D}_{\rm HOT} ({\bf k})$ around the $\Gamma=(0,0,0)$ point, we arrive at the effective low-energy model for HOT phases
\begin{eqnarray}~\label{Eq:Masterhamiltonian}
\hat{h}^{{\rm 3D},\Gamma}_{\rm HOT} ({\bf k}) &=& v \sum^2_{j=1} k_j \; \gamma_j + \left[-\frac{t_z}{2} k^2_z + \bar{\Delta} + t_0 \frac{k^2_\perp}{2}\right] \; \gamma_3 \nonumber \\
&+& \frac{1}{2}\; \Delta_1 \; (k^2_x-k^2_y) \; \gamma_4 + \Delta_2 \; k_x k_y \; \gamma_5,
\end{eqnarray} 
where $\bar{\Delta}=t_z-m_z$, $k_\perp=(k^2_x+k^2_y)^{1/2}$, and $v=t$ bears the dimension of (Fermi) velocity. For $\Delta_1=0=\Delta_2$, the system respectively describes a TDSM and a three-dimensional quantum spin Hall insulator (3D-QSHI) for $\bar{\Delta}>0 $ and $\bar{\Delta}<0$. In the former phase, the  Dirac points are located at ${\bf k}=(0,0,\pm k^{\ast}_z)$, where $k^{\ast}_z=\sqrt{2 \bar{\Delta}/t_z}$. The transition between these two phases takes place at $\bar{\Delta}=0$, where the low-energy density of states scales as $|E|^{3/2}$~\cite{roy-slager-juricic}. Both TDSM and 3D-QSHI can be constructed by stacking two-dimensional quantum spin Hall insulators (2D-QSHIs) in momentum space along the $k_z$ direction, and the corresponding band inversion takes place at $k^{\ast}_\perp=[(t_z k^2_z -2 \bar{\Delta})/t_0]^{1/2}$. Hence, for $t_0,t_z>0$, the band inversion occurs for any $k_z$ when $\bar{\Delta}<0$. By contrast, when $\bar{\Delta}>0$ it takes place only for $k_z>k^\ast_z$ and $k_z<-k^\ast_z$, and the two-dimensional insulators occupying the region $|k_z| \leq k^\ast_z$ are topologically trivial.

Both TDSM and 3D-QSHI support two copies of Fermi arc surface states, the loci of the zero-energy modes, associated with the one-dimensional counter-propagating edge states for opposite spin projections of the underlying 2D-QSHIs. Therefore, in the 3D-QSHI phase the Fermi arcs exist for any $k_z$, while two disjoint pieces of arcs are found for $k_z>k^\ast_z$ and $k_z<-k^\ast_z$ in the TDSM phase. In real space, such Fermi arc states occupy two-dimensional surfaces in the $xz$ and $yz$ planes, and hence are characterized by codimension $d_c=1$~\cite{supplementary}. Thus, the TDSM and 3D-QSHI represent first order topological phases.

Turning on $\hat{h}^{\bf k}_{1}$ (with $\Delta_2=0$) generates a mass for the one-dimensional edge states of the underlying 2D-QSHIs. This term changes sign under a discrete four-fold rotation and assumes the profile of a domain wall. Then four states at precisely zero energy appear at $\left( \pm \frac{L}{2}, \pm \frac{L}{2} \right)$ (the corners of the system), according to a generalized Jackiw-Rebbi index theorem~\cite{jackiw-rebbi}, for $|k_z|> k^{\ast}_z$ when $\bar{\Delta}>0$ or any $k_z$ when $\bar{\Delta}<0$. Here, $L$ is the linear dimension of the system in the $x$ and $y$ directions. The collection of such corner localized zero-energy modes ultimately constitutes four copies of one-dimensional \emph{hinge} states, extended along the $z$ direction, implying a second order topological phase. Even when $\Delta_2$ is finite, the hinge modes persist at half-filling [evidenced by numerically diagonalizing the tight-binding model of Eq.~(\ref{Eq:LatticeHamilToy}) on a cubic lattice], but \emph{slightly} away from zero energy due to the loss of the spectral symmetry with open boundaries~\cite{supplementary}. We continue to find similar hinge states even when the crystal is cleaved in a different orientation, such that four corners appear at $\left( \pm\frac{L}{2},0 \right)$ and $\left( 0,\pm \frac{L}{2} \right)$. But, the roles of $\Delta_1$ and $\Delta_2$ are now reversed. The phase diagram of the toy model is similar to the one shown in Fig.~\ref{Fig:Phasediagram}, but (a) the red dot represents a gapless phase with $|E|^{3/2}$ density of states at low energies and (b) the insulating phase is devoid of edge states. A different tight-binding model, obtained by taking $S_j \leftrightarrow C_j$ for $j=1,2$ and $t \to -t$ in $\hat{h}^{\rm 3D}_{\rm HOT} ({\bf k})$ [see Eq.~(\ref{Eq:LatticeHamilToy})] leads to the same low-energy model in Eq.~(\ref{Eq:Masterhamiltonian}), when expanded around $\left( \frac{\pi}{2},\frac{\pi}{2},0 \right)$. Within the framework of this lattice model $(k^2_x-k^2_y)$ and $(k_x k_y)$ arise from $(\sin k_x - \sin k_y)$ and $(\cos k_x \cos k_y)$, respectively. This, however, does not alter our conclusions. Next we proceed to demonstrate the emergence of similar HOT phases for spin-$\frac{3}{2}$ Luttinger fermions in the presence of external strain.

\emph{Luttinger model}. The Hamiltonian describing a biquadratic touching of Kramers degenerate valence and conduction bands of spin-3/2 fermions is~\cite{luttinger, murakami-nagaosa-zhang} 
\begin{eqnarray}~\label{Eq:LuttingerHamiltonian}
\hat{h}_{\rm L} ({\bf k})=- \left[\frac{1}{2 m_1} \sum^{3}_{j=1} d_j ({\bf k}) \Gamma_j + \frac{1}{2 m_2} \sum^{5}_{j=4} d_j ({\bf k}) \Gamma_j \right],
\end{eqnarray} 
where ${\boldsymbol \Gamma}$s are four-dimensional mutually anticommuting Hermitian matrices, satisfying the anticommuting Clifford algebra $\{ \Gamma_j, \Gamma_k \}=2 \delta_{jk}$, for $j,k=1,\cdots, 5$, and 
\begin{eqnarray}
d_{1}({\bf k})&=&\sqrt{3} k_y k_z, 
d_{2}({\bf k})=\sqrt{3} k_x k_z, 
d_{3}({\bf k})=\sqrt{3} k_y k_x, \nonumber \\
d_{4}({\bf k})&=&\frac{\sqrt{3}}{2} (k^2_x -k^2_y), 
d_{5}({\bf k})=\frac{1}{2} (2 k^2_z-k^2_x -k^2_y).
\end{eqnarray}  
Momenta ${\bf k}$ are measured from the band touching point, located at ${\bf k}= 0$. For the sake of simplicity, we neglect particle-hole asymmetry, and set $m_1=m_2=m$ (say), even though in a cubic environment $m_1 \neq m_2$ in general, and assume $m>0$. If we organize the four-component spinor according to $\Psi^\top_{\bf k}= ( c_{{\bf k},\frac{3}{2}}, c_{{\bf k},\frac{1}{2}}, c_{{\bf k},-\frac{1}{2}}, c_{{\bf k},-\frac{3}{2}})$, where $c_{{\bf k},m_s}$ is the fermion annihilation operator with momentum ${\bf k}$ and spin projection $m_s$, then $\Gamma_1=\kappa_3 \sigma_2, \Gamma_2=\kappa_3 \sigma_1, \Gamma_3=\kappa_2 \sigma_0, \Gamma_4=\kappa_1 \sigma_0, \Gamma_5=\kappa_3 \sigma_3$. The Pauli matrices $\{ \kappa_\mu\}$ and $\{ \sigma_\mu \}$ for $\mu=0,\cdots, 3$ respectively operate on the doublet of sign [${\rm sgn}(m_s)=\pm$] and magnitude [$|m_s| \in \{ 1/2,3/2 \}$] of the spin projection $m_s$. The $\Gamma$ matrices also provide a basis for a symmetric traceless tensor operator formed from bilinear products of spin-$\frac{3}{2}$ matrices~\cite{supplementary}.

The energy spectra of two-fold degenerate conduction and valence bands are, respectively, $\pm |{\bf k}|^2/(2m)$. While such band touching is protected by cubic symmetry, the Kramers degeneracy of the bands is ensured by  time-reversal (${\mathcal T}$) and parity (${\mathcal P}$) or inversion symmetries. Specifically, under the reversal of time ${\bf k} \to -{\bf k}$ and $\Psi_{\bf k} \to \Gamma_1 \Gamma_3 \Psi_{-{\bf k}}$, and ${\mathcal T}=\Gamma_1 \Gamma_3 \; {\mathcal K}$, where ${\mathcal K}$ is the complex conjugation, yielding ${\mathcal T}^2=-1$. Under spatial inversion, ${\bf k} \to -{\bf k}$ and $\Psi_{\bf k} \to \Psi_{-{\bf k}}$.

External strain modulates this unconventional band touching. While respecting the ${\mathcal P}$ and ${\mathcal T}$ symmetries (thus maintaining the Kramers degeneracy of the bands), it reduces the cubic symmetry of the system, and depending on its nature (compressive or tensile) destabilizes the biquadratic band touching in favor of an insulating or a gapless phase. In practice, strain is applied along high symmetry, such as the $C_{4v}$ and $C_{3v}$ axes, directions. The following discussion is geared toward unveiling the topological properties of such time-reversal invariant strain-engineered phases of spin-3/2 fermions.

\emph{$E_g$ strain}. The coupling of a generic strain, transforming under the $E_g$ representation, with spin-3/2 Luttinger fermions is captured by the Hamiltonian $\hat{h}_{E_g}= \Delta \left( \sin \phi \; \Gamma_4 + \cos \phi \; \Gamma_5 \right)$, where $|\Delta|$ is its amplitude and $\phi$ is the internal phase~\cite{roy-ghorashi-foster-nevidomskyy, szabo-moessner-roy}. An external strain applied along the $z$ direction is parametrized by $\phi=0$ (tensile) or $\pi$ (compressive). On the other hand, the tensile and compressive strain along the $x \; (y)$ direction respectively correspond to $\phi=2\pi/3 \; (4\pi/3)$ and $5\pi/3 \; (\pi/3)$. If we denote the Hamiltonian operator associated with strain along $j$ direction as $\hat{h}^{j}_{E_g}$, then 
\begin{eqnarray}~\label{Eq:Egstraingeneral}
\hat{h}^{z}_{E_g} &=& |\Delta| \Gamma_5 \; {\rm sgn}(\Delta), \:
\hat{h}^{x}_{E_g}= \frac{|\Delta|}{2} \left[ \sqrt{3} \Gamma_4 - \Gamma_5 \right] {\rm sgn}(\Delta), \nonumber \\
\hat{h}^{y}_{E_g} &=& - \frac{|\Delta|}{2} \left[ \sqrt{3} \Gamma_4 + \Gamma_5 \right] {\rm sgn}(\Delta).
\end{eqnarray}  
In this notation, the tensile and compressive strain correspond to ${\rm sgn}(\Delta)= +$ and $-$, respectively.

We first focus on strain applied along the $z$ direction, for which the total Hamiltonian reads
\begin{eqnarray}~\label{Eq:EgztotalHamil}
\hat{h}^{z,{\rm t}}_{E_g} &=& v{_z}  [k_x \Gamma_2 + k_y \Gamma_1] + \left[-\frac{k^2_z}{2 m} + |\Delta| {\rm sgn}(\Delta) +\frac{k^2_\perp}{4 m}\right] \Gamma_5 \nonumber \\
&-& \frac{1}{2m} \left[\frac{\sqrt{3}}{2}(k^2_x-k^2_y) \Gamma_4+ \sqrt{3} k_x k_y \Gamma_3 \right],   
\end{eqnarray} 
and can be identified with $\hat{h}^{\rm 3D}_{\rm HOT} ({\bf k})$, see Eq.~(\ref{Eq:Masterhamiltonian}), with
$$v=v_{_z}, \; 
t_z=2t_0=\frac{1}{m}, \;  
\bar{\Delta}=\Delta, \;
\Delta_1=\Delta_2=-\frac{\sqrt{3}}{2 m},$$
where $v_z=\sqrt{3}k_z/(2m)$, and $\gamma_1=\Gamma_2, \gamma_2=\Gamma_1, \gamma_3=\Gamma_5,\gamma_4=\Gamma_4,\gamma_5=\Gamma_5$.
The only (and important) difference is that $v_z$ vanishes for $k_z=0$. Otherwise, for any finite $k_z$, the emergent topology of strained spin-3/2 Luttinger fermions follows the previous discussion. For example, tensile strain produces a TDSM with Dirac points located at $(0,0,\pm k_z^0)$, where $k^0_z=\sqrt{2m |\Delta|}$, while compressive strain yields a topological insulator~\cite{szabo-moessner-roy, herbut-janssen, liu-strain, goswami-roy-dassarma}. We next focus on the $k_z=0$ plane to fully characterize the emergent topology and the resulting boundary modes of these two phases.

In the $k_z=0$ plane, $\hat{h}^{z,{\rm t}}_{E_g}$ from Eq.~(\ref{Eq:EgztotalHamil}) can be cast (after suitable unitary rotation~\cite{supplementary, kharitonov}) in a block-diagonal form $H_+ \oplus H_-$, where for $\sigma=\pm$
\begin{equation}
H_\sigma = \left[ |\Delta| \; {\rm sgn}(\Delta) +\frac{k^2_\perp}{4 m}\right] \sigma \tau_3 -\frac{1}{2m} \left[d_4({\bf k}) \tau_1 + d_3({\bf k}) \tau_2 \right],
\end{equation}  
since it involves three mutually anti-commuting $\Gamma$ matrices. For ${\rm sgn}(\Delta)=+$, the $k_z=0$ plane hosts a trivial insulator (devoid of band inversion), supporting no boundary mode. Therefore, the TDSM hosts one-dimensional hinge modes with $d_c=2$ only for $|k_z|>k_z^0$, extended along the $z$ direction, and is thus of second-order. By contrast, when ${\rm sgn}(\Delta)<0$, the $k_z=0$ plane is occupied by a 2D-QSHI, composed of two superimposed copies of quantum anomalous Hall insulators of Chern numbers $C=\pm 2$ for opposite spin projections ($\sigma=\pm$). It accommodates two copies of counter-propagating edge modes for opposite spin projections that reside in the $xz$ and $yz$ planes. For any finite $k_z$, the insulating phase supports only corner localized zero modes, ultimately constituting four hinge states along the $z$ direction. Hence, the resulting three-dimensional topological insulator hosts both one-dimensional hinge modes (with $d_c=2$) and two-dimensional edge modes ($d_c=1$), and thus supports \emph{mixed} topological boundary modes. The resulting phase diagram of strained spin-3/2 Luttinger fermions is schematically shown in Fig.~\ref{Fig:Phasediagram}.

In the rest of the Rapid Communication, we show that when external strain is applied along a certain high symmetry direction, the effective single-particle Hamiltonian always takes the form of $\hat{h}^{z,{\rm t}}_{E_g}$ [see Eq.~(\ref{Eq:EgztotalHamil})], after suitable relabeling of the momentum axes and redefinitions of mutually anticommuting $\Gamma$ matrices. Hence, the above discussion is sufficient to capture the emergent topology of strain-engineered phases in Luttinger systems, at least when strain is applied along high-symmetry directions. To explicitly demonstrate this in our present framework, we continue with the $E_g$ strain, and construct the equivalent of the Hamiltonian in Eq.~(\ref{Eq:EgztotalHamil}) for strain along the (symmetry equivalent) $x$ direction. This proceeds by introducing new sets of momenta $q_x=k_z, q_y=k_y, q_z=k_x$, and four-component Hermitian matrices $\bar{\Gamma}_1=\Gamma_3$, $\bar{\Gamma}_2=\Gamma_2$, $\bar{\Gamma}_3=\Gamma_1, \bar{\Gamma}_4=( \Gamma_4+\sqrt{3} \; \Gamma_5 )/2$ and $\bar{\Gamma}_5=( \sqrt{3} \; \Gamma_4- \Gamma_5)/2$, such that they also satisfy the anticommuting Clifford algebra $\{ \bar{\Gamma}_j, \bar{\Gamma}_k \}=2 \delta_{jk}$, for $j,k=1,\cdots,5$ and $q_z \parallel [100]$. We then find
\begin{equation}
\hat{h}_{\rm L} ({\bf k}) + \hat{h}^{x}_{E_g} \equiv -\frac{1}{2m} \sum^{5}_{j=1} d_j ({\bf q}) \bar{\Gamma}_j + |\Delta| \; \bar{\Gamma}_5 \; {\rm sgn}(\Delta).
\end{equation} 
Therefore, the resulting TDSM phase [${\rm sgn}(\Delta)=+$] supports only four hinge modes along the $x$ direction, while the insulator [${\rm sgn}(\Delta)=-$] in addition accommodates edge mode residing on the $xy$ and $xz$ planes. Following similar steps~\cite{supplementary}, analogous results follow for strain and the concomitant hinge modes in the $y$ direction, alongside the edge modes in the $xy$ and $yz$ planes. Next we proceed to discuss the effects of external strains, now applied along the (symmetry inequivalent) $C_{3v}$ or body diagonal or $\langle 111 \rangle$ directions.

\emph{$T_{2g}$ strain}. The effective single-particle Hamiltonian, capturing the coupling between an external strain applied along a body-diagonal or $C_{3v}$ axis of a cubic system and spin-3/2 Luttinger fermions is given by  
\begin{equation}~\label{Eq:T2gnematic}
\hat{h}_{T_{2g}}=\frac{|\Delta|}{\sqrt{3}} \: \sum^3_{j=1} \; {\rm sgn} (\Delta_j) \; \Gamma_j, 
\end{equation}
where $|\Delta|=\sqrt{\Delta^2_1 + \Delta^2_2 + \Delta^2_3}$ and $\Delta_{1,2,3}$ are the components of the strain along one of the [111] directions in the $x,y,z$ axes, respectively. In this notation, tensile and compressive strain respectively correspond to ${\rm sgn}(\Delta)=\pm$, where ${\rm sgn} (\Delta)=\prod^3_{j=1} {\rm sgn} (\Delta_j)$. To demonstrate the effects of a $T_{2g}$ strain, we  align it along one of the specific body-diagonal directions, the $[1,1,1]$, direction, for which either (a) ${\rm sgn} (\Delta_j)=+$ (tensile) or (b) ${\rm sgn} (\Delta_j)=-$ (compressive), for $j=1,2,3$. This yields (a) a pair of Dirac points along the [111] body diagonal at ${\bf k}=\pm k^0_{[111]}(1,1,1)$, where $k^0_{[111]}=\sqrt{2 m |\Delta|/3}$ or (b) an insulator.

To address the emergent topology of these two phases, we introduce a new set of momenta with 
$q_z$ $\parallel$ [111]
\begin{eqnarray}
q_x=\frac{k_x-k_y}{\sqrt{2}}, \;
q_y=\frac{k_x+k_y-2 k_z}{\sqrt{6}},\;
q_z=\frac{k_x+k_y+k_z}{\sqrt{3}},\nonumber 
\end{eqnarray} 
and define a new set of $\bar{\Gamma}$ matrices satisfyig the Clifford algebra $\{ \bar{\Gamma}_j, \bar{\Gamma}_k \}=2\delta_{jk}$, for $j,k=1\cdots 5$, 
\begin{eqnarray}
\bar{\Gamma}_1 &=& -\left[ \frac{1}{3\sqrt{2}} \left( \Gamma_1 + \Gamma_2 -2 \Gamma_3 \right) + \sqrt{\frac{2}{3}} \; \Gamma_5 \right], \nonumber \\
\bar{\Gamma}_2  &=&  -\frac{\Gamma_1- \Gamma_2-2 \Gamma_4 }{\sqrt{6}}, \;
\bar{\Gamma}_3  =   \frac{\Gamma_1 -\Gamma_2 + \Gamma_4 }{\sqrt{3}}, \nonumber \\
\bar{\Gamma}_4 &=&  \frac{\Gamma_1+\Gamma_2-2 \Gamma_3-\sqrt{3} \Gamma_5}{3}, \;
\bar{\Gamma}_5  =   \frac{\Gamma_1 + \Gamma_2 + \Gamma_3}{\sqrt{3}}. \nonumber 
\end{eqnarray}
The effective single particle Hamiltonian reads 
\begin{equation}
\hat{h}_{\rm L} ({\bf k}) + \frac{|\Delta|}{\sqrt{3}} \; {\rm sgn} (\Delta) \sum^3_{j=1} \Gamma_j 
\equiv
\hat{h}_{\rm L} ({\bf q}) + |\Delta| \bar{\Gamma}_5 \; {\rm sgn}(\Delta),
\end{equation} 
similar to $\hat{h}^{z,{\rm t}}_{E_g}$, see Eq.~(\ref{Eq:EgztotalHamil}). Hence, the resulting HOT TDSM phase hosts four copies of hinge modes along the [111] direction. Besides such hinge modes, the strain engineered insulator supports edge modes localized on two-dimensional surfaces perpendicular to the [111] direction. One arrives at analogous conclusions when the external strain is applied along one of the other three body-diagonal directions, namely $[1\bar{1}\bar{1}]$, $[\bar{1}1\bar{1}]$ and $[\bar{1}\bar{1}1]$, directions~\cite{supplementary}.

\emph{Summary and discussion}. We show that a family of electronic systems, described by a collection of cubic spin-3/2 fermions, which in the normal state display a biquadratic band touching, can harbor HOT phases (such as TDSM and insulator), when they are subjected to external strains, applied along certain high symmetry directions (such as $\langle 100 \rangle$ and $\langle 111 \rangle$) directions, see Fig.~\ref{Fig:Phasediagram}. Some well known members of this family are HgTe~\cite{hgte}, gray-Sn~\cite{gray-sn-1, gray-sn-2}, 227 pyrochlore iridates~\cite{Savrasov, Exp:Nakatsuji-1,Exp:Nakatsuji-2} and half-Heusler compounds~\cite{Exp:cava, Exp:felser, binghai}. Since the inversion symmetry breaking in HgTe and half-Heuslers is very weak~\cite{hgte,Exp:cava, Exp:felser, binghai, ISB-comment}, in the presence of moderate strain they will nonetheless support HOT Weyl semimetals or inversion-odd insulators, hosting Kramers non-degenerate hinge modes in the strain direction. It will be interesting to further explore the fate of strain engineered topological phases in noncentrosymmetric Luttinger materials~\cite{Rappe, Soluyanov}. It should also be worthwhile investigating potential HOT phases in Dirac materials such as A$_3$Bi (with A=Na,K,Rb)~\cite{xi-dai-na3bi} and $\beta$-CuI~\cite{thomale-Dirac}, where a cubic higher gradient term (transforming under the $E_u$ representation of D$_{3d}$ group) can be mimicked by taking $\hat{h}^{\bf k}_1 \to C_z \hat{h}^{\bf k}_1$ [see Eq.~(\ref{Eq:LatticeHamilToy})]. Such a cubic term is absent in the Luttinger Hamiltonian, as it does not transform under any irreducible representation of the cubic $O_h$ (227 pyrochlore iridate, gray-Sn), zinc blende (HgTe) or tetrahedral $T_d$ (half-Heusler) point group.

Experimentally, tunable compressive and tensile strains can be applied using piezoelectric-based uniaxial stress apparatus~\cite{Hicks-2014} (by gluing the sample between the jaws of the uniaxial pressure device~\cite{Clifford-CPFS}) or on thin films by growing them on a substrate with slightly different lattice constant (see Ref.~\cite{nakatsuji-thinfilm}). Although high-quality 227 pyrochlore iridate crystals tend to be small, methods to apply uniaxial stress to miniaturized samples are under active development~\cite{clifford-comment}. While both insulator and TDSM support four hinge modes in the strain direction, the former one in addition accommodates edge modes on two orthogonal planes, which can be detected from angle-resolved photoemission spectroscopy (ARPES) and scanning tunneling microscope (STM) measurements. Recently engineered TDSM in strained HgTe can be a good platform to test our predictions using STM~\cite{molenkamp-HgTe}. The application of strain along one of the $C_{4v}$ (such as $[001]$) is promising in gray-Sn, whereas our findings in the presence of $T_{2g}$ strain can be relevant for 227 pyrochlore iridates (due to the natural growth of pyrochlore crystals in the $\langle 111 \rangle$ direction)~\cite{nakatsuji-thinfilm}. Even though pyrochlore iridates~\cite{iridates-globalPD} and half-Heuslers~\cite{heuslers-globalPD} support various magnetic and/or superconducting phases at lowest temperature, corresponding ordering temperature can be suppressed significantly by applying hydrostatic and chemical pressures~\cite{ueda-pressure-1, ueda-pressure-2}. Therefore, our proposed strain engineered HOT phases in the world of spin-3/2 Luttinger fermions can be germane above the ordering temperatures in strongly correlated systems and should motivate future experimental works.

\emph{Acknowledgments}. This work was in part supported by DFG through SFB 1143 (project-ID 247310070) and \emph{ct.qmat} (EXC 2147, project-ID 39085490). We are thankful to Clifford Hicks and Andrew Mackenzie for valuable discussion. B.R. was partially supported by the startup Grant from Lehigh University.

\end{document}